\documentclass[amssymb,aps,prb,letterpaper,twocolumn,showpacs]{revtex4-1}
\usepackage[utf8]{inputenc}
\usepackage{graphicx}
\usepackage{amsmath,amsfonts,amssymb}
\usepackage{physics}
\usepackage{multirow}
\usepackage{comment}
\usepackage{bbm}
\usepackage{psfrag}

\newcommand{\identity}{{\mathbbm{1}}}
\usepackage[dvipsnames]{xcolor}

\begin{document}
\title{Numerical evidences of a universal critical behavior
of 2D and 3D random quantum clock and Potts models}
\author{Valentin Anfray}
\affiliation{Universit\'e de Lyon, Universit\'e Claude Bernard Lyon 1, CNRS,
Institut Lumi\`ere Mati\`ere, F-69622 Villeurbanne, France}
\author{Christophe Chatelain}
\affiliation{Universit\'e de Lorraine, CNRS, LPCT, F-54000 Nancy, France}
\date{\today}

\begin{abstract}
The random quantum $q$-state clock and Potts models are studied in
2 and 3 dimensions. The existence of Griffiths phases is tested in
the 2D case with $q=6$ by sampling the integrated probability distribution
of local susceptibilities of the equivalent McCoy-Wu 3D classical models
with Monte Carlo simulations. No Griffiths phase is found for the
clock model. In contrast, numerical evidences of the existence of
Griffiths phases in the random Potts model are given and the
Finite Size effects are analyzed. The critical point of the random
quantum clock model is then studied by Strong-Disorder Renormalization Group.
Despite a chaotic behavior of the Renormalization-Group flow at weak
disorder, evidences are given that this critical behavior is governed
by the same Infinite-Disorder Fixed Point as the Potts model,
independently from the number of states $q$.
\end{abstract}
\maketitle

\section{Introduction}
Universality is a cornerstone of the theory of critical phenomena.
At the vicinity of a continuous phase transition, the critical exponents
appearing in the algebraic behavior of thermodynamic quantities are
independent from the microscopic details of the system. The Renormalization
Group provides a theoretical foundation to this universality~\cite{Cardy}.
The symmetry of the Hamiltonian that is spontaneously broken during the
phase transition participates to determine the universality class.
Universality classes may be drastically affected by the introduction
of disorder in the system.
The $q$-state Potts model, whose Hamiltonian is invariant under the
permutations of the $q$ states, is a well-known example of this principle.
In two dimensions, the classical Potts model undergoes a second-order
phase transition when $q\le 4$ with critical exponents depending on the
number of states $q$~\cite{Wu}. As shown by Harris,
the critical behavior is unchanged in presence of randomness only if the
specific heat of the pure model diverges at most logarithmically~\cite{Harris}.
In contrast, when the specific heat exponent $\alpha$ is positive,
randomness induces a new critical behavior. This the case for the
2D classical Potts model when $2<q\le 4$~\cite{Ludwig,Dotsenko,Chatelain1}.
Moreover, discontinuous phase transitions are smoothed by randomness.
It was proved that, in two dimensions, an infinitesimal amount of disorder
is sufficient to make the transition continuous~\cite{Aizenmann}.
Such a situation is observed for the 2D Potts model with $q>4$.
The new critical behavior is again $q$-dependent~\cite{Jacobsen1,Chatelain2}.
\\

Surprisingly, a totally different picture emerged in random quantum systems.
In the models considered in the literature, the critical behavior turns
out to be independent of the spontaneously-broken symmetry of the Hamiltonian.
The random Ising chain in a transverse field has been extensively studied
by Fisher~\cite{Fisher1,Fisher2,Monthus1,Monthus2} using a real-space
Renormalization Group technique, termed as Strong-Disorder Renormalization Group
(SDRG)~\cite{MaDasgupta}. In contrast to the pure quantum Ising chain in
a transverse field, the dynamics is activated, i.e. at the critical point the
gap $\Delta E$ vanishes with the lattice size $L$ as $\Delta E\sim e^{-aL^\psi}$
instead of the usual law $\Delta E\sim L^{-z}$. The dynamical exponent $z$ is
therefore infinite. The exponent $\psi$ takes the value $1/2$ because
the logarithm of the renormalized couplings generated along the SDRG
flow are uncorrelated random variables. The magnetic critical exponent is
predicted to be equal to the golden ratio $(1+\sqrt 5)/2$. On both
sides of the critical point lie two Griffiths phases where the dynamics
is dominated by rare macroscopic regions with strong fluctuations of
randomness. The dynamical exponent is equal to 1 at the boundary of these
phases and diverges as the critical point is approached~\cite{Igloi1}.
The $q$-state random quantum Potts chain was also studied by
SDRG~\cite{Senthil}. Note that the Ising model in a transverse field
is equivalent to the $q=2$ Potts model. The analysis of the RG flow
equations showed that the number of states $q$ is irrelevant. Therefore, the
critical behavior of the random Potts chain is governed by the same Infinite
Disorder Fixed Point (IDFP) as the Ising chain for any number of states
$q$. The fact that the symmetry-group of the Hamiltonian is $q$-dependent
does not play any role. DMRG simulations confirmed this result~\cite{Carlon}.
The $q$-state random quantum clock chain was also considered. In contrast
to the Potts model, the Hamiltonian is invariant under circular permutations
of the $q$ states. SDRG also predicts that the critical behavior of the
random clock chain is described by the same $q$-independent IDFP as the
Ising and Potts chains at strong enough disorder~\cite{Senthil}. This
result was also confirmed by DMRG simulations~\cite{Carlon2}. Finally, the
Ashkin-Teller model, corresponding to two coupled Ising chains, was studied
using a numerical implementation of SDRG. Again, along the three transitions
lines, the critical behavior is governed by the same IDFP as the Ising chain.
Only at the tricritical point where these lines merge, a different IDFP
was identified~\cite{Vojta1,Vojta2,Chatelain3}.
\\

An interesting question in this context is whether this super-universality
is specific to one-dimensional quantum systems or also exists in higher
dimensions. The random Ising model in a transverse field was studied
in dimensions $d=2,3$ and 4 using a clever numerical implementation
of SDRG~\cite{Kovacs1,Kovacs2,Kovacs3}. It turns out that the critical
exponents depends on the dimension $d$. In particular, the exponent
$\psi$ takes values smaller than $1/2$, indicating that the logarithm
of the renormalized couplings are now correlated. Recently, we applied
the same algorithm to the 2D and 3D random Potts models and provided
numerical evidences that the critical behavior is the same as the Ising model
at the same dimension $d$ for all the number of states $q$
considered~\cite{Anfray}.
\\

In this work, the random $q$-state clock and Potts models are considered.
In the first section, the 3D classical analogue of the 2D random quantum
6-state clock and Potts models are studied by Monte Carlo simulations.
While a phase transition is clearly seen, no Griffiths phase could be
observed for the clock model. In the second section, the critical behavior
of the 2D and 3D random quantum $q$-state clock models is determined by
numerically applying the SDRG technique for several numbers of states
$q$. It is then compared to the Potts and Ising models. Conclusions follow.

\section{The classical 3D 6-state Potts and clock models}
The classical $q$-state Potts model is defined by the Hamiltonian
   \begin{equation}
    -\beta{\cal H}=\sum_{(i,j)} J_{ij}\delta_{\sigma_i,\sigma_j},
    \quad \sigma_i=0,\ldots,q-1
    \label{HPotts}
    \end{equation}
while for the clock model
    \begin{equation}
    -\beta{\cal H}=\sum_{(i,j)} J_{ij}\cos\Big({2\pi\over q}(\sigma_i-\sigma_j)\Big),
    \quad \sigma_i=0,\ldots,q-1.
    \label{HClock}
    \end{equation}
where $\beta=1/k_BT$. The temperature is absorbed into the definition of the
couplings. The sum extends over all pairs $(i,j)$ of nearest-neighboring sites
on the lattice. In this section, the two models are considered in the case $q=6$
on a cubic lattice whose sizes are denoted $L_\parallel$, $L_\perp$ and $L_\perp$.
In the pure case, i.e. $J_{ij}=J$, both models undergo a ferromagnetic-paramagnetic
phase transition. The latter is strongly first order for the Potts model
and continuous in the universality class of the 3D XY model for the clock
model~\cite{Scholten,Miyashita,Hove}. In both cases, the order parameter is
the magnetization density ${1\over N}|\sum_i m_i|$ where the local magnetization
reads
    \begin{equation}
    m_i={q\delta_{\sigma_i,0}-1\over q-1}
    \label{MagnetizationPotts}
    \end{equation}
for the Potts model, and
    \begin{equation}
    m_i=e^{{2\imath\pi\over q}\sigma_i}
    \label{MagnetizationClock}
    \end{equation}
for the clock model. In the following, we are interested in the random Potts and
clock models with a disorder that is infinitely correlated in one direction.
The couplings $J_{ij}$ between sites $i$ and $j$ were chosen to depend
only on the coordinates $y$ and $z$ of site $i$, i.e. they are
translation-invariant in the $x$ direction. The three couplings between site $i$
and its neighbors on the right, at the bottom and at the back, are identical.
The couplings are i.i.d random variables with the binary distribution
    \begin{equation}
    \wp(J_{ij})={1\over 2}\big(\delta(J_{ij}-J)+\delta(J_{ij}-rJ)\big).
    \end{equation}
This configuration is one of the possible generalization of the McCoy-Wu model
to 3D~\cite{McCoyWu}. Since the specific heat exponent of the pure 3D XY
model is slightly negative~\cite{Gottlob,Campostrini,Xu}, or equivalently the
correlation length exponent $\nu$ is slightly larger than $2/d$, the pure fixed
point of the clock model is expected to be stable upon the introduction of
homogeneous disorder~\cite{Harris}. However, $\nu<2/(d-1)=1$ so the
infinitely-correlated disorder that is studied in this paper is expected to be
relevant. For the 6-state Potts model, the first-order phase transition
of the pure model is expected to be rounded, even with an infinitesimal
amount of disorder~\cite{Aizenman}.

\begin{figure}
    \psfrag{J1}[Bl][Bl][1][1]{$J$}
    \psfrag{Chiperp}[Bl][Bl][1][1]{$\chi_\perp$}
    \psfrag{Chiparallel}[Bl][Bl][1][1]{$\chi_\parallel$}
    \centering
    \includegraphics[width=0.47\textwidth]{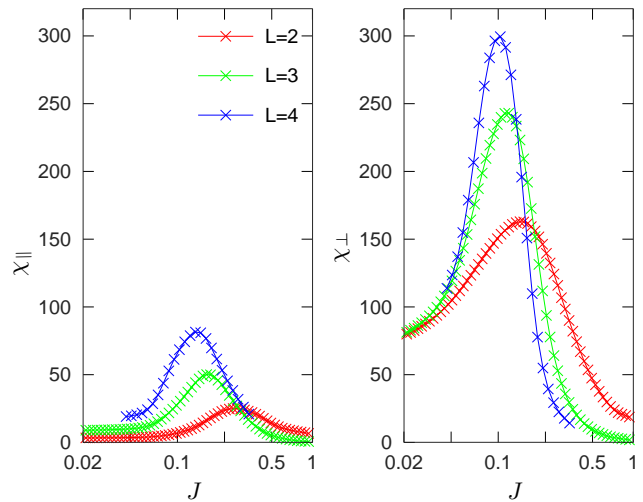}
    \caption{Longitudinal and transverse susceptibilities $\chi_\parallel$
    (left) and $\chi_\perp$ (right) of the 3D 6-state random clock model versus
    the weakest exchange coupling $J$. The different curves correspond to
    different transverse lattice sizes $L_\perp$. The longitudinal lattice
    size is $L_\parallel=64$.}
    \label{fig1}
\end{figure}

For practical reasons, we studied the clock model using the Wolff Monte
Carlo algorithm~\cite{Wolff} and the Potts model using the Swendsen-Wang
algorithm~\cite{Swendsen}. Both algorithms are equivalent in terms of
convergence. In the following, one Monte Carlo Step (MCS) corresponds
to $N=L_\parallel L_\perp^2$ spin flips on average. We first considered small
systems, $L_\perp=2,3,4$ and $L_\parallel=64$, so that the average over
all disorder configurations could be performed exactly. For $L_\perp=4$,
the number of disorder configurations is $2^{L_\perp^2}=65536$.
The reason for this choice is that we observed large deviations of the
averages when computed over a smaller number of disorder configurations.
The location of the phase transition can be estimated as the position of
the maximum of the average magnetic susceptibility
    \begin{equation}
    \bar\chi=N\big[\overline{\langle m^2\rangle}
    -\overline{\langle m\rangle^2}\big].
    \label{Chi}
    \end{equation}
where the brackets $\langle X\rangle$ denote the average over thermal
fluctuations while the above line $\overline{X}$ indicates the average
over disorder. The system being strongly anisotropic, it is useful to define
the two susceptibilities:  
    \begin{eqnarray}
    \bar\chi_\parallel&=&{1\over NL_\parallel}\sum_{y,z}\overline{\Big\langle
    \Big|\sum_x m_{(x,y,z)}\Big|^2\Big\rangle},
    \nonumber\\
    \bar\chi_\perp&=&{1\over NL_\perp^2}\sum_{x}\overline{\Big\langle\Big|
    \sum_{y,z} m_{(x,y,z)}\Big|^2\Big\rangle}.
    \label{Chi2}
    \end{eqnarray}
where $m_{(x,y,z)}$ denotes the local magnetization density on the site whose
Cartesian coordinates on the cubic lattice are $(x\ y\ z)$. For both the Potts
and clock models, a single peak is observed for both susceptibilities
$\bar\chi_\parallel$ and $\bar\chi_\perp$, indicating the existence of a single
phase transition. As can be seen on Fig.~\ref{fig1}
for the clock model with $r=10$, the shift of the location of the maximum of
the peak is larger for $\chi_\parallel$ than for $\chi_\perp$. The same
picture is qualitatively observed with a weaker disorder $r=5$, apart from the
fact that the peaks of $\chi_\parallel$ and $\chi_\perp$ are almost of same
height. The truncation of the average to a small fraction of all disorder
configurations leading to strong deviations, the critical behavior at the
transition could be not studied by Monte Carlo simulations. In the next
section, the Strong Disorder Renormalization Group will be applied to the
quantum equivalent of the models. Nevertheless, the existence of Griffiths
phases around the critical point can be investigated.
\\

In a random system, large local fluctuations of the couplings allow for
the existence of macroscopic ordered (resp. unordered) regions while
the rest of the system is still in the paramagnetic (resp. ferromagnetic)
phase~\cite{Griffiths}. The average magnetization is a non-analytic
function of the magnetic field in both the ordered and disordered Griffiths
phases that surrounds the critical point. As a consequence, the magnetic
susceptibility diverges in the whole Griffiths phases.
These singularities are too weak to be observed in classical systems
with homogeneous disorder but not in the McCoy-Wu model, for which
the couplings are infinitely correlated in one direction. Due to
the limitation to small lattice sizes, the divergence of the
susceptibility in a finite range of couplings is not observed on
Fig.~\ref{fig1}. However, the Griffiths phases also manifest themselves
by an algebraic decay of the probability distribution of local linear
and non-linear susceptibilities~\cite{Rieger}. The advantage is that
the method does not require an exact average over all disorder configurations.
The local susceptibility is defined as
    \begin{equation}
    \chi_{\rm loc}=L_\parallel\big[\langle m_1^2\rangle-\langle m_1\rangle^2\big]
    \end{equation}
and the non-linear susceptibility (or Binder cumulant) as
    \begin{eqnarray}
    U_{\rm loc}&=&L_\parallel^2\big[\langle m_1^4\rangle
    -4\langle m_1\rangle\langle m_1^3\rangle
    +12\langle m_1\rangle^2\langle m_1^2\rangle\big. \nonumber\\
    &&\hskip 2cm
    -3\langle m_1^2\rangle^2-6\langle m_1\rangle^4\big]
    \end{eqnarray}
where the magnetization density $m_1={1\over L_\parallel}\sum_x m_{(x,1,1)}$
is computed over the first row of the lattice only. The local susceptibility
$\chi_{\rm loc}$ is related to the correlation length in the longitudinal direction:
     \begin{eqnarray}
     \chi_{\rm loc}&=&{1\over L_\parallel}\sum_{x,x'}\big[
     \langle m_{(x,1,1)}m_{(x',1,1)}\rangle
     -\langle m_{(x,1,1)}\rangle^2\big]\rangle      \nonumber\\
     &\simeq& {1\over L_\parallel}\int_0^{L_\parallel}
     e^{-|x-x'|/\xi_\parallel}dxdx'                 \nonumber\\
     &\simeq& 2\xi_\parallel\quad\quad\quad (\xi_\parallel\ll L_\parallel)
     \label{RelChiXi}
     \end{eqnarray}
The disordered Griffiths phase is due to clusters of strong couplings $rJ$
that will order before the rest of the system. The probability of such a cluster
of characteristic size $\ell$ decays exponentially fast as
    \begin{equation}
    \wp(\ell^{d_\perp})\sim p^{\ell^{d_\perp}}
    \label{ProbaCluster}
    \end{equation}
where $p$ is the probability of a strong coupling (chosen to be $p=1/2$ in this
work) and $d_\perp=d-1$. When the temperature $T$ is between the critical
temperature of the strong and weak couplings, this cluster tends to order
ferromagnetically because the free energy gain in volume compensates the loss
at its boundaries with the paramagnetic phase:
    \begin{equation}
    \Delta F=-\ell^{d_\perp}L_\parallel\big[f_{\rm o}-f_{\rm d}\big]
    +\ell^{d_\perp-1}L_\parallel\sigma_{\rm o,d}>0
    \label{Bilan}
    \end{equation}
where $f_{\rm o}(rJ)$ and $f_{\rm d}(J)$ are the free energy densities
of the ordered and disordered homogeneous phases for strong and weak couplings
respectively. $\sigma_{o,d}(J,rJ)$ is the surface tension between
ordered and disordered phases. It follows from Eq.~\ref{Bilan} that there exists a
minimal size $\ell_{\rm min}=\sigma_{o,d}/[f_{\rm o}-f_{\rm d}]$ for a ferromagnetic
cluster to be stable. Moreover, the long-range ordering of the cluster in
the longitudinal direction may be reduced by the coexistence of different
ferromagnetic phases separated by domain walls whose free energy cost is
    \begin{equation}
    \Delta F_{\rm DW}=\sigma_{o,o}\ell^{d_\perp}
    \end{equation}
per domain wall, where $\sigma_{o,o}$ is the surface tension between two
ordered phases. The average size of a magnetic domain is therefore
related to the size of the cluster as
    \begin{equation}
    \xi_\parallel\sim e^{\beta\sigma_{o,o}\ell^{d_\perp}}
    \end{equation}
Since $\xi_\parallel\simeq\chi_{\rm loc}$ according to Eq.~\ref{RelChiXi},
and since the probability of a cluster of strong couplings of characteristic
size $\ell$ decays algebraically with $\ell^d_\perp$ as Eq.~\ref{ProbaCluster},
it follows that $\wp(\ln\chi_{\rm loc})\simeq \wp(\ln\xi_\parallel)\sim
\wp(\ell^d)$ and therefore~\cite{Rieger}
    \begin{equation}
    \ln\wp(\ln\chi_{\rm loc})={\rm Cst}-{d_\perp\over z}\ln\chi_{\rm loc}
    \label{ProbaChi}
    \end{equation}
where $d_\perp/z=-\ln p{k_BT\over\sigma_{o,o}}$.
The same behavior is expected for the Binder cumulant but with a coefficient
$d_\perp/3z$ instead of $d_\perp/z$. This prediction has been exploited to
estimate the dynamical exponent $z$ of the transverse-field Ising
spin-glass~\cite{Rieger,Guo} and of the random Ising ferromagnet~\cite{Kawashima}.
Note finally that the average susceptibility $\overline{\chi_{\rm loc}}
=\int \chi_{\rm loc}\wp(\chi_{\rm loc})d\chi_{\rm loc}$ diverges when
$z>1/d_\perp$.
\\

\begin{figure}
    \psfrag{1-F(chi)}[Bl][Bl][1][1]{$1-F(\chi_{\rm loc})$}
    \psfrag{F(U)}[Bl][Bl][1][1]{$F(U_{\rm loc})$}
    \psfrag{chi}[Bl][Bl][1][1]{$\chi_{\rm loc}$}
    \psfrag{U}[Bl][Bl][1][1]{$U_{\rm loc}$}
    \centering
    \includegraphics[width=0.47\textwidth]{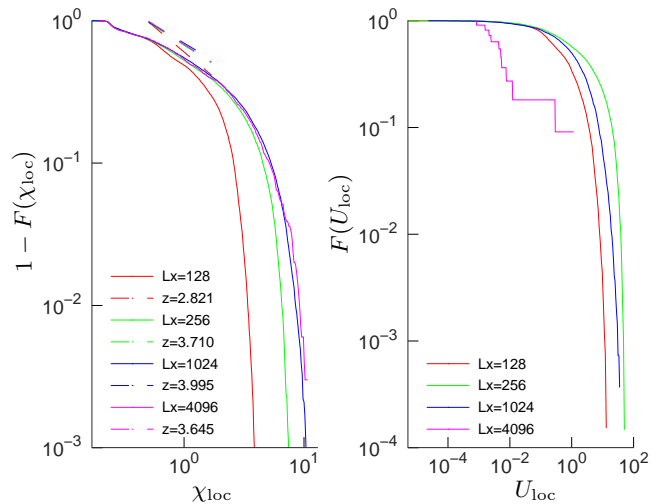}
    \caption{Integrated probability distribution of the local
    susceptibility $\chi_{\rm loc}$ (left) and of the Binder cumulant
    $U_{\rm loc}$ (right) of the 6-state random clock model at
    $J=0.0772$ and $r=10$. The different curves correspond to
    different longitudinal lattice sizes $L_\parallel$. The transverse
    lattice size is $L_\perp=12$.}
    \label{figHisto1}
\end{figure}

\begin{figure}
    \psfrag{1-F(chi)}[Bl][Bl][1][1]{$1-F(\chi_{\rm loc})$}
    \psfrag{F(U)}[Bl][Bl][1][1]{$F(U_{\rm loc})$}
    \psfrag{chi}[Bl][Bl][1][1]{$\chi_{\rm loc}$}
    \psfrag{U}[Bl][Bl][1][1]{$U_{\rm loc}$}
    \centering
    \includegraphics[width=0.47\textwidth]{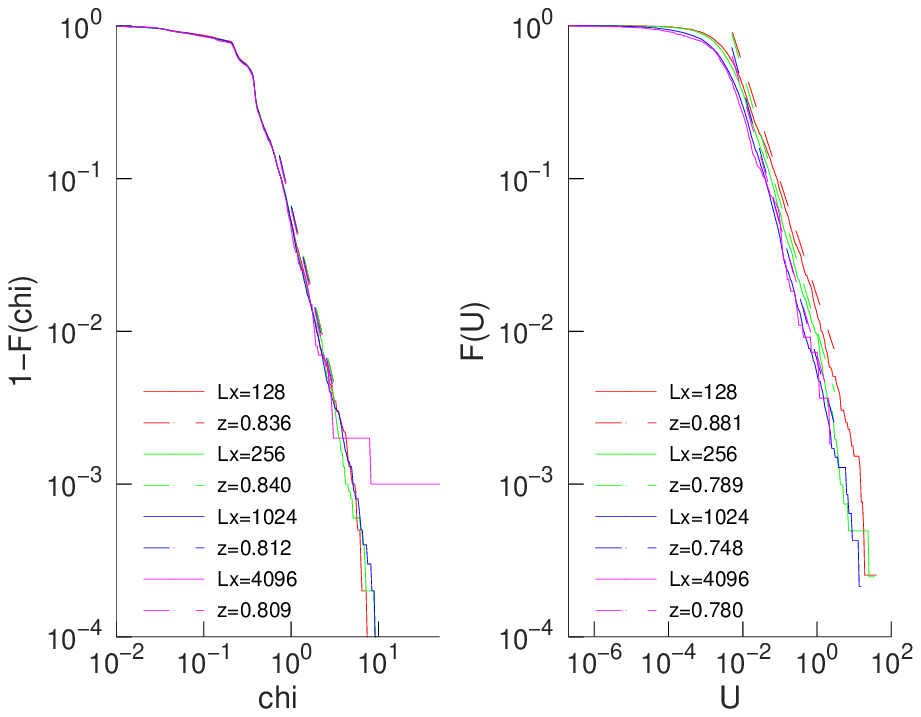}
    \caption{Integrated probability distribution of the local
    susceptibility $\chi_{\rm loc}$ (left) and of the Binder cumulant
    $U_{\rm loc}$ (right) of the 6-state random Potts model at
    $J=0.1156$ and $r=10$. The different curves correspond to
    different longitudinal lattice sizes $L_\parallel$. The transverse
    lattice size is $L_\perp=12$.}
    \label{figHisto4}
\end{figure}

We performed Monte Carlo simulations of the random clock and Potts models with
lattice sizes $L_\perp=12$ and $L_\parallel=128,256,1024$ and 4096. The system
was thermalized with 1000 MC iterations and thermal averages were then
computed over 10000 MC iterations. 10000 random configurations were sampled
for $L_\parallel\le 1024$ and 1000 for $L_\parallel=4096$. We computed the
integrated probability distributions $1-F(\chi_{\rm loc})$ and $F(U_{\rm loc})$
that are expected to display the same algebraic decay in Griffiths phases as
$\wp(\ln\chi_{\rm loc})$ and $\wp(\ln U_{\rm loc})$. Examples of integrated
probability distributions of both $\chi_{\rm loc}$ and $U_{\rm loc}$ are
plotted in figures~\ref{figHisto1} to \ref{figHisto4}.
For the clock model, we could not find any coupling $J$ for which the
integrated probability distributions of both $\chi_{\rm loc}$ and $U_{\rm loc}$
display a clear algebraic decay. The only coupling for which $1-F(\chi_{\rm loc})$
can reasonably be fitted with a power-law over a small range of
susceptibilities $\chi_{\rm loc}$ is shown on Fig.~\ref{figHisto1}.
The fit is plotted as a dashed line which was slightly shifted vertically
to be visible. The associated dynamical exponent is
$z\simeq 3.8$ at large lattice size $L_\parallel$. However, the integrated
probability distribution $F(U_{\rm loc})$ does not display any algebraic
decay at the same coupling. Our conclusion is that there is no
Griffiths phase in the phase diagram of the random clock model.
\\

\begin{figure}
    \psfrag{1-F(chi)}[Bl][Bl][1][1]{$1-F(\chi_{\rm loc})$}
    \psfrag{F(U)}[Bl][Bl][1][1]{$F(U_{\rm loc})$}
    \psfrag{chi}[Bl][Bl][1][1]{$\chi_{\rm loc}$}
    \psfrag{U}[Bl][Bl][1][1]{$U_{\rm loc}$}
    \centering
    \includegraphics[width=0.47\textwidth]{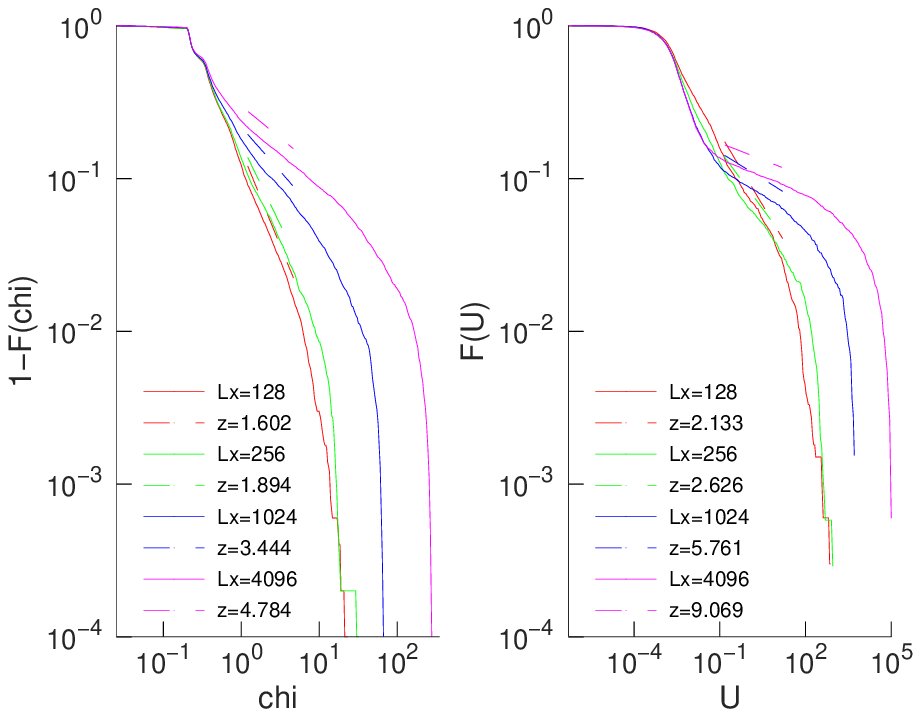}
    \caption{Integrated probability distribution of the local
    susceptibility $\chi_{\rm loc}$ (left) and of the Binder cumulant
    $U_{\rm loc}$ (right) of the 6-state random Potts model at
    $J=0.1037$ and $r=10$. The different curves correspond to
    different longitudinal lattice sizes $L_\parallel$. The transverse
    lattice size is $L_\perp=12$.}
    \label{figHisto3}
\end{figure}

\begin{figure}
    \psfrag{1-F(chi)}[Bl][Bl][1][1]{$1-F(\chi_{\rm loc})$}
    \psfrag{F(U)}[Bl][Bl][1][1]{$F(U_{\rm loc})$}
    \psfrag{chi}[Bl][Bl][1][1]{$\chi_{\rm loc}$}
    \psfrag{U}[Bl][Bl][1][1]{$U_{\rm loc}$}
    \centering
    \includegraphics[width=0.47\textwidth]{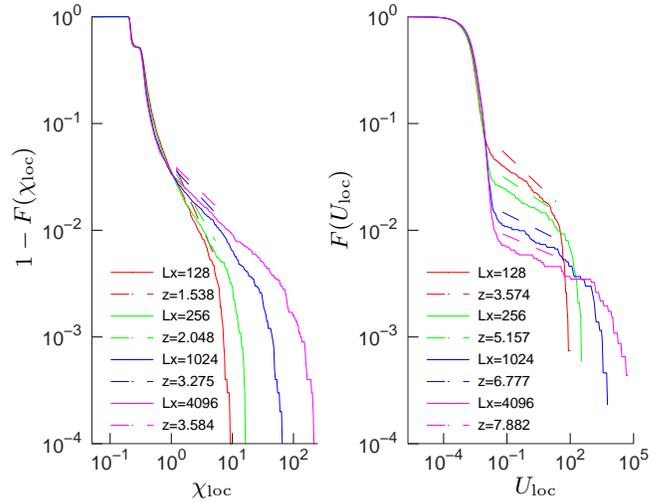}
    \caption{Integrated probability distribution of the local
    susceptibility $\chi_{\rm loc}$ (left) and of the Binder cumulant
    $U_{\rm loc}$ (right) of the 6-state random Potts model at
    $J=0.0951$ and $r=10$. The different curves correspond to
    different longitudinal lattice sizes $L_\parallel$. The transverse
    lattice size is $L_\perp=12$.}
    \label{figHisto2}
\end{figure}

The situation is different for the Potts model. At $J=0.1156$ (Fig.~\ref{figHisto4}),
both the integrated probability distributions $1-F(\chi_{\rm loc})$ and
$F(U_{\rm loc})$ display an algebraic decay over several decades. The slopes
of the curves in a log-log scale give estimates of the dynamical
exponent around $z\simeq 0.8$. The latter being larger than $1/d_\perp=1/2$,
the system is in a Griffiths phase. At smaller couplings, closer to the
critical point, the dynamical exponent is expected to be larger. However,
an algebraic decay is still observed on Fig.~\ref{figHisto3} ($J=0.1037$)
and \ref{figHisto2} ($J=0.0951$) but over a smaller range of susceptibilities.
Various Finite-Size effects can be invoked to explain the shape of the curves.
A first regime, where the decay is slow and not algebraic, is observed at
small susceptibilities $\chi_{\rm loc}$ and $U_{\rm loc}$. This regime is due
to clusters that are too small to satisfy Eq.~\ref{Bilan}, i.e.
$\ell<\ell_{\rm min}$. At intermediate susceptibilities, an algebraic decay
is observed but with a strong dependence on the longitudinal lattice size
$L_\parallel$. This dependence can be attributed to Eq.~\ref{RelChiXi}.
For large correlation lengths $\xi_\parallel$, one should replace
Eq.~\ref{RelChiXi} by (with Periodic Boundary Conditions)
    \begin{equation}
    \chi_{\rm loc}\simeq 2\xi_\parallel
    \big(1-e^{-L_\parallel/\xi_\parallel}\big)
    \end{equation}
which leads to a probability distribution $\wp(\ln\chi_{\rm loc})
=\wp(\ln\xi_\parallel){d\ln\xi_\parallel\over d\ln\chi_{\rm loc}}$
that depends on $L_\parallel$, in contrast to Eq.~\ref{ProbaChi}.
Finally, at large susceptibilities, the probability distributions fall down
much faster than a power-law. This is due to the fact that the size $\ell$
of the clusters is bounded by the transverse lattice size $L_\perp$. As a
consequence, the susceptibility $\chi_{\rm loc}$ cannot be larger than
$e^{\beta\sigma_{o,o}L_\perp^{d_\perp}}$. In conclusion, we infer that the
random Potts model is in a Griffiths phase for $J=0.0951$ and $0.1037$ but
that Finite-Size effects are strong and explain why the algebraic decay
is observed only over a smaller range of susceptibilities. The dynamical
exponents that can be estimated using Eq.~\ref{ProbaChi} are, as expected,
large. However, they strongly depends on $L_\parallel$. These estimates
should be considered with caution because a more accurate estimate would
require to use the exact $L_\parallel$-dependent expression of
$\wp(\ln\chi_{\rm loc})$ and not Eq.~\ref{ProbaChi} that is valid only
in the limit $\xi_\parallel\ll L_\parallel$.

\section{Critical behavior of 2D and 3D random quantum clock models}
In this section, the critical behavior of the 2D and 3D random quantum $q$-state
clock model is studied by Strong Disorder Renormalization Group. Note that
the 2D and 3D random quantum $q$-state Potts models have been considered in
Ref.~\cite{Anfray}. In the extreme anisotropic limit, the transfer matrix of
the classical $q$-state clock model (Eq.~\ref{HClock}) in dimension $d+1$
is equivalent to the imaginary-time evolution operator of the quantum clock
model in dimension $d$ whose Hamiltonian is
    \begin{equation}
    H=-\sum_{(i,j)} J_{ij}\big(\Omega_i\Omega_j^++\Omega_j\Omega_i^+\big)
    -\sum_i h_i\big(M_i+M_i^+\big)
    \label{HQuantique}
    \end{equation}
where the sum extends over nearest neighbors on the lattice. The
operator $\Omega_i$ acts only on the spin on site $i$, i.e.
$\Omega_i=\identity^{\otimes i-1}\otimes\Omega\otimes\identity^{\otimes N-i}$
where $N$ is the number of sites of the lattice. The matrix $\Omega$ is a
$q\times q$ diagonal matrix whose elements are $\omega^n$ with
$\omega=e^{2i\pi/q}$. Similarly, $M_i=\identity^{\otimes i-1}\otimes M
\otimes\identity^{\otimes N-i}$ and $M$ is the $q\times q$ matrix whose
only non-zero elements are equal to 1 and are on the diagonal above
the main diagonal. The couplings $J_{ij}$ and $h_i$ are positive, random
and uncorrelated.

\subsection{SDRG decimation rules and chaoticity of the RG flow}
\begin{figure}
    \psfrag{h/J}[Bl][Bl][1][1]{$h/J$}
    \psfrag{heff}[Bl][Bl][1][1]{$\tilde h/J$}
    \centering
    \includegraphics[width=0.37\textwidth]{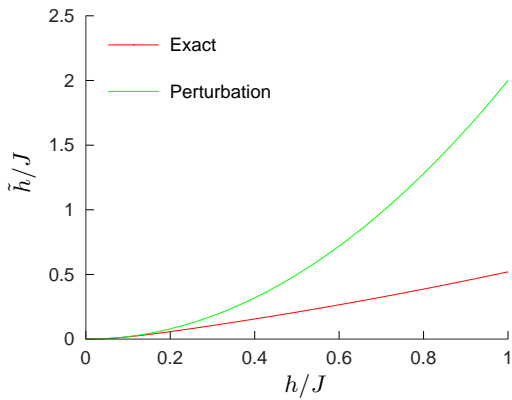}
    \caption{Effective transverse field $\tilde h$ versus $h/J$ ($h_1=h_2=h$)
    computed either by second-order perturbation theory (\ref{SDRGRule1})
    or by exact diagonalization of the two-site Hamiltonian
    $-J(\Omega_1\Omega_2^++\Omega_2\Omega_1)^+-h(M_1+M_1^+)-h(M_2+M_2^+)$
    for the 6-state clock model.
    In the second case, the effective transverse field is extracted from
    the gap $\Delta E$ between the first excited state and the ground state
    as $\tilde h=\Delta E/2\kappa$.}
    \label{fig3}
\end{figure}

The SDRG decimation rules are easily derived for the $q$-state clock
model~\cite{Senthil}. When the largest coupling is $\Omega=J_{ij}$, the
Hamiltonian (\ref{HQuantique}) is projected onto the ground state of the
local Hamiltonian $-J_{ij}\big(\Omega_i\Omega_j^++\Omega_j\Omega_i^+\big)$.
Since the latter is $q$-fold degenerated, the two spins on sites $i$ and
$j$ are replaced by an effective $q$-state macro-spin. Second-order
perturbation theory shows that this macro-spin is coupled to an
effective transverse field
    \begin{equation}
    \tilde h={h_ih_j\over\kappa J_{ij}}
    \label{SDRGRule1}
    \end{equation}
where $\kappa=1-\cos 2\pi/q$ for $q>2$ and to all the neighbors of
sites $i$ and $j$ by a coupling $J_{ik}+J_{jk}$. When the largest coupling
is $\Omega=h_i$, the spin is frozen in the ground state of the local
Hamiltonian $-h_i\big(M_i+M_i^+\big)$. An effective coupling between
any pairs $(k,l)$ of neighboring sites coupled to $i$ is
induced at second-order perturbation theory:
    \begin{equation}
    \tilde J_{kl}=J_{kl}+{J_{ki}J_{il}\over \kappa h_i}.
    \label{SDRGRule2}
    \end{equation}
For the $q$-state Potts model, the same decimation rules are obtained
but with $\kappa=q/2$. This small difference has however important
consequences. For the Potts model, the effective coupling is always
smaller than $\Omega$. It is not always the case for the clock model
since $\kappa<1$ for $q>4$. When a bond $\Omega=J_{ij}$ is decimated,
an effective transverse field $\tilde h$ larger than $\Omega$
can be generated when the two transverse fields $h_i$ and $h_j$
are large, more precisely when $h_ih_j>\kappa\Omega$ although
$h_i,h_j<\Omega$.
This large induced transverse field $\tilde h$ is not physical and comes
from the fact that the renormalized couplings are computed using
second-order perturbation theory. At higher orders, the renormalized
couplings are expected to remain smaller than $\Omega$. As can be seen
on Fig.~\ref{fig3}, the comparison with an exact diagonalization of
the 2-site Hamiltonian shows that the use of second-order perturbation
theory is justified when $h_1,h_2\lesssim 0.2J_{ij}$ for the 6-state clock model
while the effective transverse field becomes larger than $\Omega=J_{ij}$
when $h_1,h_2\simeq 0.7J_{ij}$.
The consequences of this are severe only far from the IDFP.
As the IDFP is approached, the probability distribution of the couplings
becomes broader and broader. As a consequence, the probability that
$h_ih_j>\kappa\Omega$, and therefore that an effective transverse field
$\tilde h>\Omega$ be induced by the decimation of $J_{ij}$, becomes
smaller and smaller.
However, numerical simulations are limited to finite systems and
therefore to a finite number of RG steps. For small systems, the
simulation may end up still relatively far from the IDFP. We observed
that renormalized couplings $\tilde h,\tilde J>\Omega$ slow down the
convergence of the RG flow towards the IDFP. In some situations, the
RG flow can even become chaotic. Note that the renormalized couplings are
proportional to $1/\kappa$ so the effect is stronger for clock models
with large numbers of states $q$.

One way to overcome the problem posed by large renormalized couplings is to
modify the decimation rules, either by including higher-orders in perturbation
or simply by using {\sl ad-hoc} effective couplings equal to (\ref{SDRGRule1})
and (\ref{SDRGRule2}) in the limit $\tilde h,\tilde J\ll \Omega$.
Another possibility is to start the simulation with broad initial
distributions of the couplings $h_i$, $J_{ij}$, close to those expected at
the IDFP. We used a SDRG algorithm similar to the one employed for the Potts
model~\cite{Anfray} but with a decimation of the global maximum and
not of the local maximum. This algorithm has the advantage to limit drastically
the number of new lattice bonds generated during the RG flow by the use of
maximum rule but it requires the effective couplings to be of the form
(\ref{SDRGRule1}) and (\ref{SDRGRule2}). As a consequence, we will avoid
large effective couplings by starting the simulations with broad coupling
distributions
        \begin{eqnarray}
        &&P_0(J_{ij})\sim J_{ij}^{1/\Delta-1},\quad (0<J_{ij}<1)\nonumber\\
        &&Q_0(h_i)\sim h_i^{1/\Delta-1},\quad (0<h_i<h_{\rm max}).
        \label{Proba2}
        \end{eqnarray}
where $\Delta$ is a free parameter controlling the broadness of the distributions
and therefore the strength of the distribution. $\theta=\ln h_{\rm max}$ plays
the role of the control parameter of the quantum phase transition. At large
(resp. small) $\ln h_{\rm max}$, the system is expected to be in the paramagnetic
phase (resp. ferromagnetic phase).

\begin{figure}
    \psfrag{h/J}[Bl][Bl][1][1]{$h/J$}
    \psfrag{heff}[Bl][Bl][1][1]{$\tilde h/J$}
    \centering
    \includegraphics[width=0.235\textwidth]{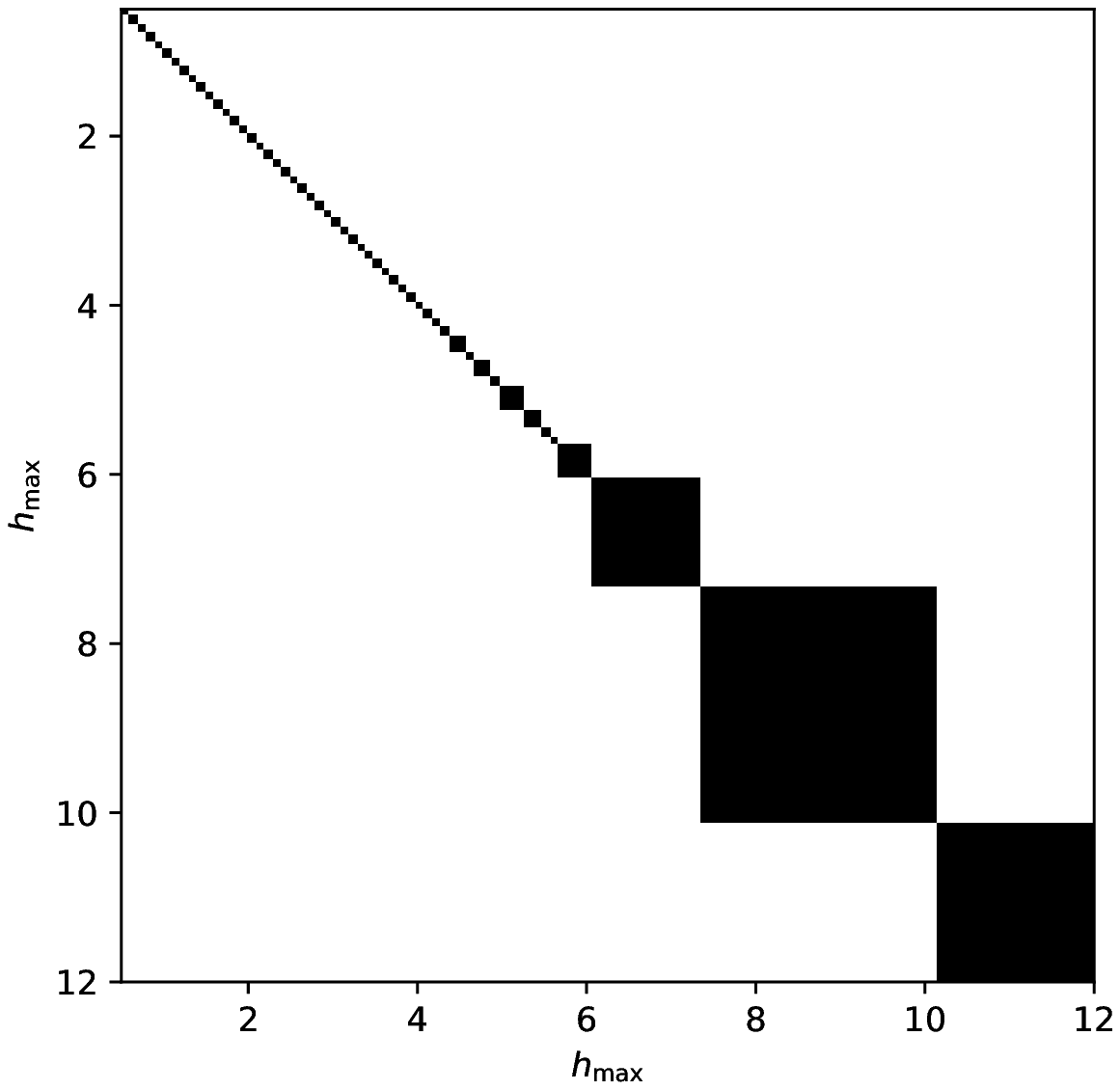}
    \includegraphics[width=0.235\textwidth]{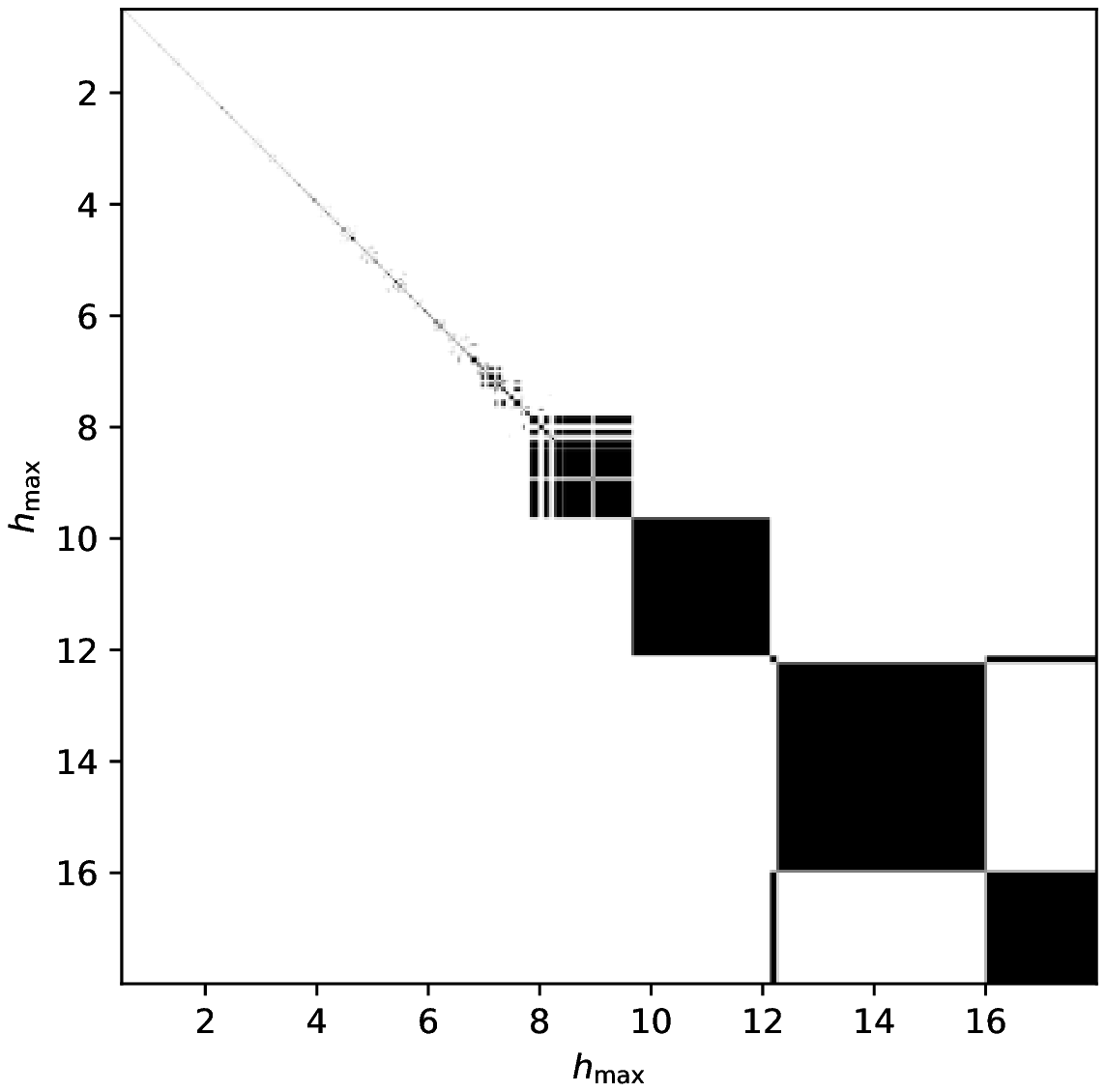}\\
    \includegraphics[width=0.235\textwidth]{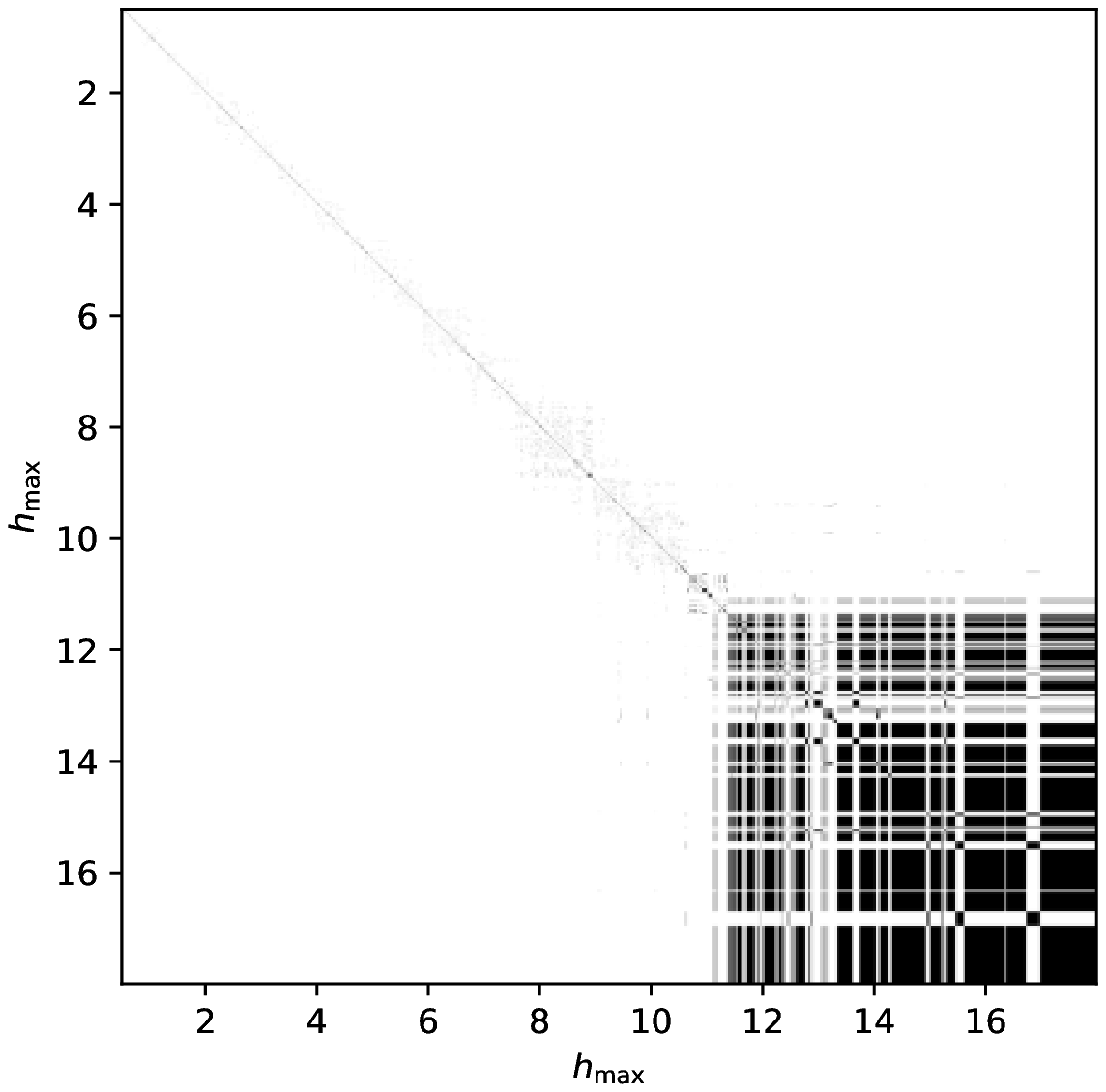}
    \includegraphics[width=0.235\textwidth]{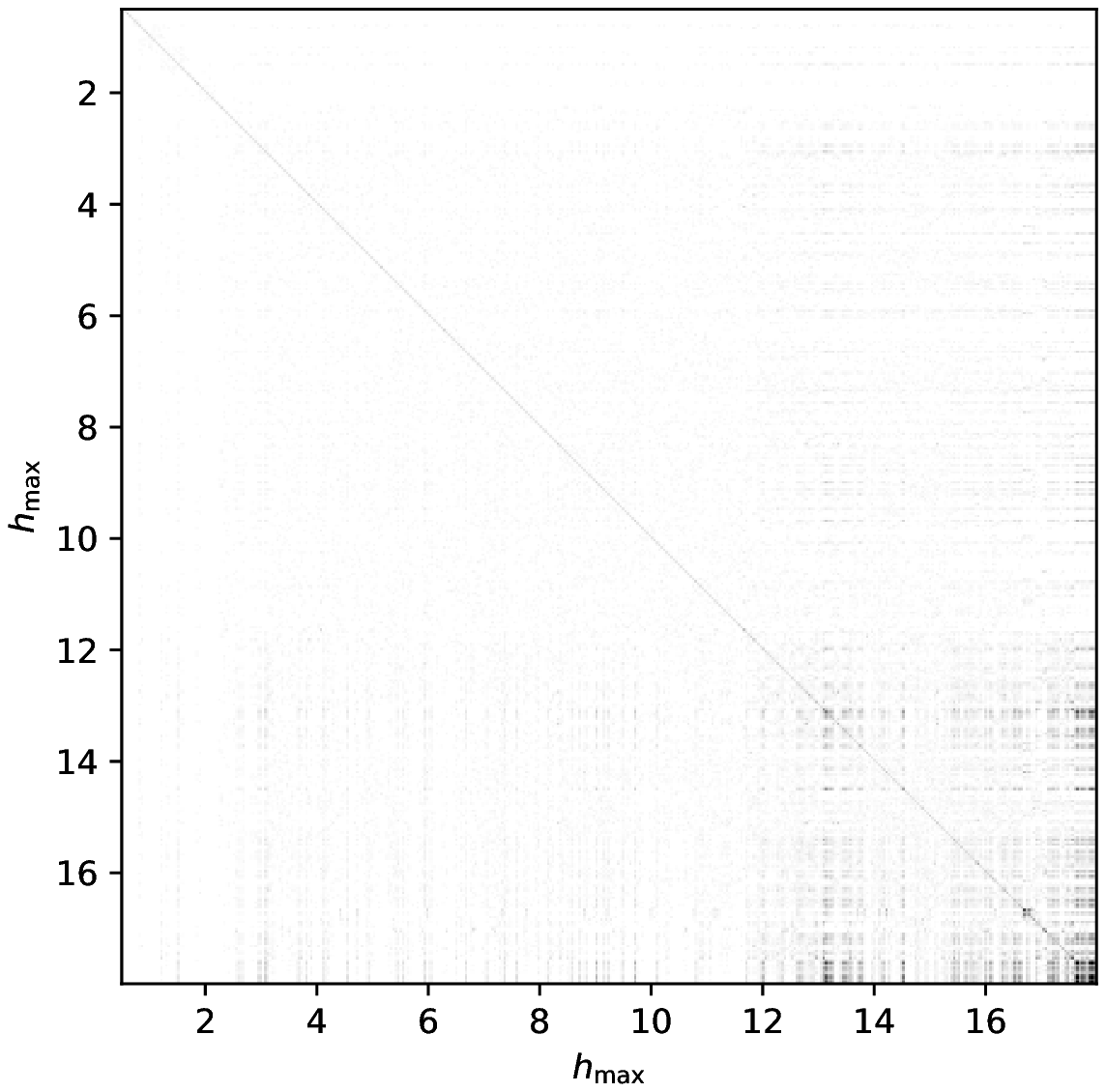}
    \caption{Recurrence matrices for a $32\times 32$ clock model with uniform
    distributions $P_0(J_{ij})$ and $Q_0(h_i)$ of the initial couplings.
    The number of states of the clock model is 2 (top left), 6 (top right),
    8 (bottom left) and 10 (bottom right).}
    \label{fig4}
\end{figure}

The chaotic nature of the RG flow manifests itself in the behavior of the average
magnetic moment $\bar\mu$ of the last cluster at the end of the RG procedure.
The latter is expected to be a smooth function of $h_{\rm max}$. However, we observed
that it is not the case for weak initial disorder or large number of states $q$.
To demonstrate the appearance of chaoticity, we applied the RG procedure to the
same disorder realization for $n$ values $h_1<h_2<\ldots <h_n$ of the maximal
transverse field $h_{\rm max}$ and measured the magnetic moment $\mu_i=\mu(h_i)$ of
the last macro-spin. A $n\times n$ recurrence matrix $R$
is constructed as $R_{ij}=1$ if $\mu_i=\mu_j$ and $R_{ij}=0$ otherwise. The matrix
is plotted on Fig.~\ref{fig4} for four different numbers of states $q$ of the clock model.
For $q=2$, the matrix is block diagonal: the magnetic moment is a monotonous non-chaotic
step function of the control parameter $h_{\rm max}$. In contrast, for $q=10$, the
magnetic moment oscillates widely with $h_{\rm max}$ and, as a consequence,
the recurrence matrix becomes noisy. Fig.~\ref{fig4} shows that
chaoticity increases with the number of states $q$. We have also considered the
recurrence matrices for a fixed number of states $q$ and an increasing
strength of disorder $\Delta$. As mentioned earlier, chaoticity decreases
with $\Delta$. In the next section, $\Delta$ is always chosen sufficiently large
to avoid, or at least strongly suppress, the chaoticity of the RG flow.

\subsection{Critical exponents at the IDFP}
We considered quantum $q$-state clock models with $q=5,8$ and $10$. We
added the cases $q=2$ and $3$ that are equivalent to the Potts models and
were already studied~\cite{Anfray}. Lattice
sizes up to $L=768$ (1536 in some cases) have been reached in 2D and $L=90$
(120 in some cases) in 3D. Averages were performed over at least $8.10^3$
disorder realizations. In the following, we will use the control parameter
$\theta=\ln h_{\rm max}$.
\\

\begin{figure}
    \centering
    \includegraphics[width=0.50\textwidth]{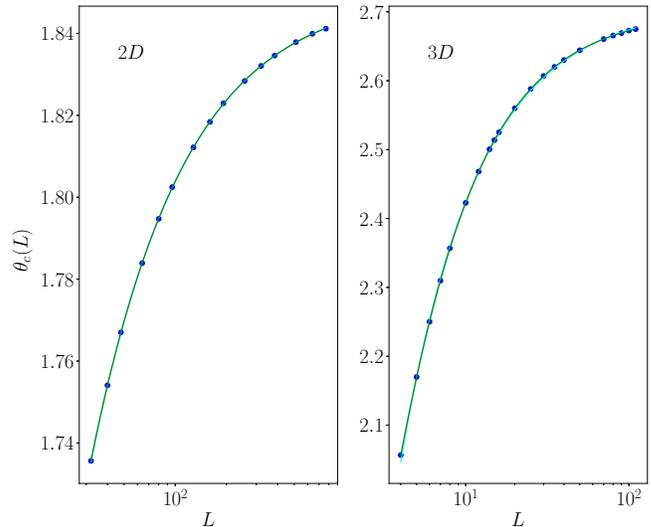}
    \caption{Average control parameter $\overline{\theta_c(L)}$ at the
    pseudo-critical critical point versus the lattice $L$ for the 2D (left)
    and 3D (right) random 10-state clock model with $\Delta=5$.
    The blue curve corresponds to a fit to Eq. (\ref{Theta1})
    while scaling corrections are taken into account in the fit plotted
    in green color. As can be seen, the two curves can hardly be
    distinguished.}
    \label{fig5}
\end{figure}

\begin{figure}
    \centering
    \includegraphics[width=0.50\textwidth]{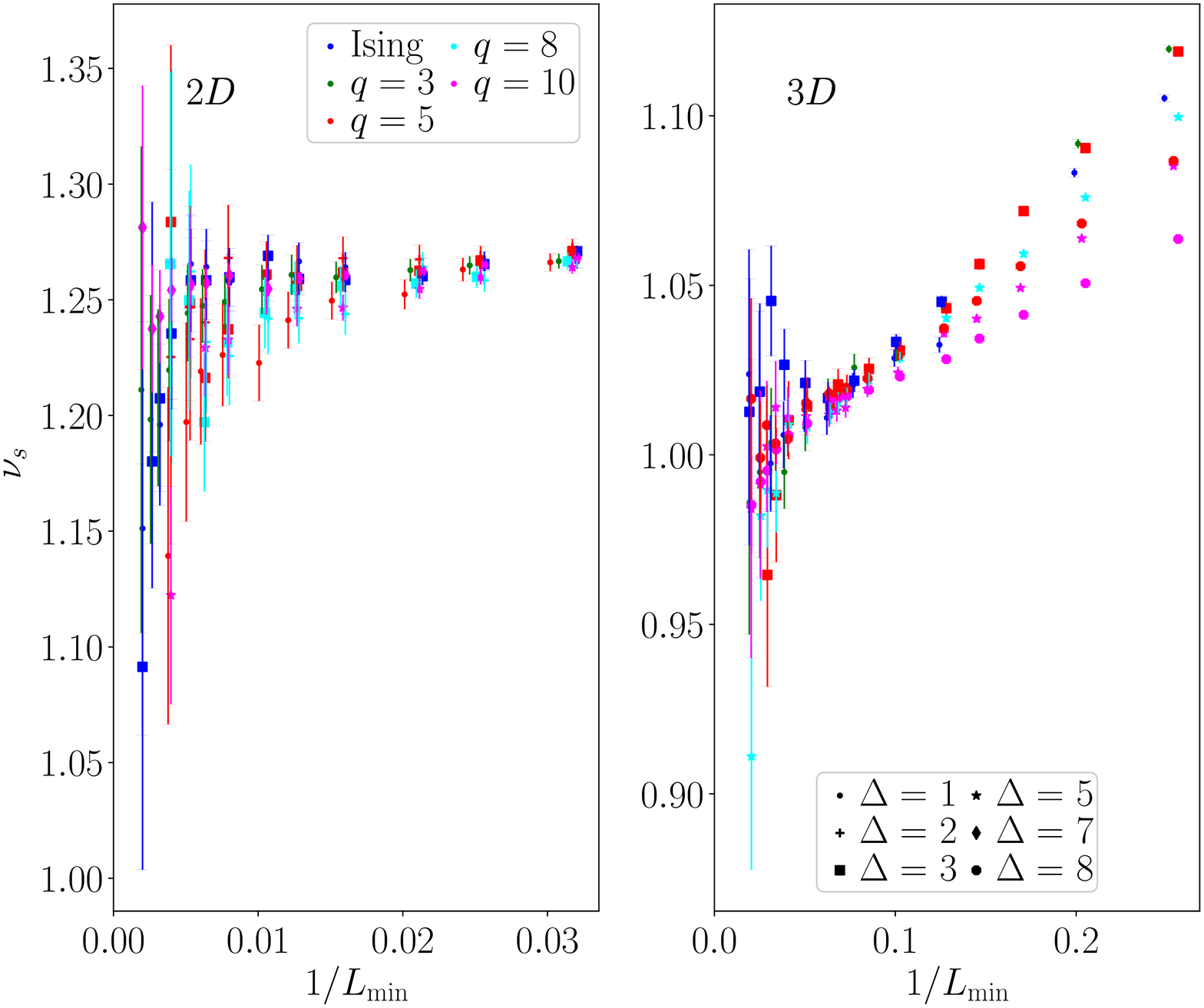}
    \caption{Critical exponent $\nu$ extracted from the finite-size behavior
    of the shift of the average control parameter $\overline{\theta_c(L)}$
    at the pseudo-critical critical point versus the inverse $1/L_{\rm min}$
    of the smallest lattice size $L_{\rm min}$ taken into account in the fit.
    The left figure corresponds to the 2D random $q$-state clock model
    and the right one to the 3D model. The color of the symbols is related to the
    number of states $q$ and their shape to the strength of the initial disorder.}
    \label{fig6}
\end{figure}

The control parameter $\theta_c(L)$ at the critical point is estimated using
the doubling method for each random sample~\cite{Kovacs1,Kovacs2}. The SDRG
procedure is applied to two replicas of the same system with the same
disorder configuration coupled together at their boundaries. The pseudo-critical
point $\theta_c(L)$ is localized iteratively by using the fact that the
magnetic moment of the last cluster of the replicated system is expected to
be twice the magnetization of a single replica in the ferromagnetic phase
but the same in the paramagnetic phase. For large lattice sizes $L$,
the average pseudo-critical control parameter $\overline{\theta_c(L)}$
displays the power-law behavior
    \begin{equation}
    |\overline{\theta_c(L)}-\theta_c|\sim L^{-1/\nu}.
    \label{Theta1}
    \end{equation}
The data are shown on Fig~\ref{fig5} in the case of the 2D and 3D random
10-state clock model. The critical exponent $\nu$ has been estimated by
a non-linear fit with the law Eq.~\ref{Theta1}. However, although small,
the existence of scaling corrections can be seen on the data, especially
in 3D. To take them into account, we restricted the fit to lattice sizes
$L>L_{\rm min}$ and repeated the operation with increasing sizes
$L_{\rm min}$. The effective exponents $\nu(L_{\rm min})$ given by these
fits are plotted versus $1/L_{\rm min}$ in Fig~.\ref{fig6}.
Most of the points are compatible within error bars and no indication
of a dependence on the number of states $q$ is observed.
Extrapolating to $L_{\rm min}\rightarrow +\infty$, the exponent can be
estimated to be $\nu\simeq 1.24(3)$ in 2D and $0.99(3)$ in 3D.
These values are in agreement with those obtained for the Potts model:
$\nu\simeq 1.25(6)$ in the 2D case and $1.01(5)$ in 3D~\cite{Anfray}.
The strong dependence of the effective exponent on $L_{\rm min}$ in 3D
suggests the existence of $q$-dependent scaling corrections. We have
also performed fits with the scaling law $|\overline{\theta_c(L)}
-\theta_c|\sim L^{-1/\nu}(1+aL^{-\omega})$ including the first scaling
correction. However, the fit becomes unstable and the error bars on $\nu$
are large in the 2D case while it gives compatibles estimates of $\nu$
for 3D clock models.

\begin{figure}
    \centering
    \includegraphics[width=0.50\textwidth]{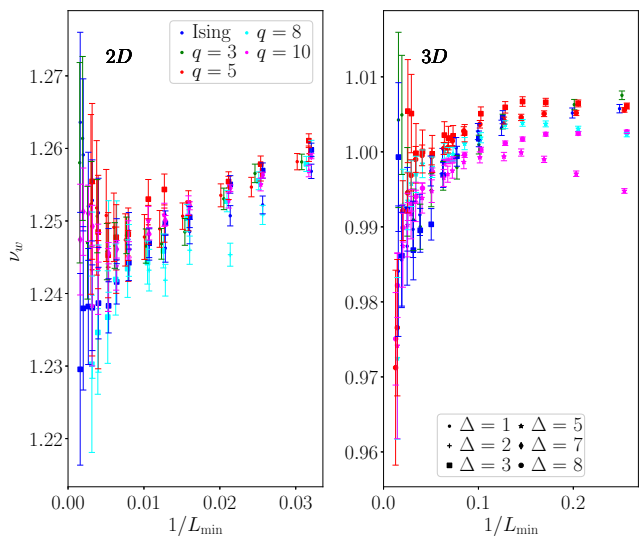}
    \caption{Critical exponent $\nu$ extracted from the finite-size behavior
    of the standard deviation $\Delta\theta_c(L)$ of the control parameter
    at the pseudo-critical critical point versus the inverse $1/L_{\rm min}$
    of the smallest lattice size $L_{\rm min}$ taken into account in the fit.
    The left figure corresponds to the 2D random $q$-state clock model
    and the right one to the 3D model. The color of the symbols is related to the
    number of states $q$ and their shape to the strength of the initial disorder.}
    \label{fig7}
\end{figure}

The critical exponent $\nu$ can also be estimated from the Finite-Size
Scaling of the standard deviation of the pseudo-critical points $\Delta\theta_c(L)
=[\overline{(\theta_c(L))^2}-\overline{\theta_c(L)}^2]^{1/2}$. The same
procedure as above is applied to take into account scaling corrections.
The effective exponents $\nu(L_{\rm min})$ are presented on Fig.~\ref{fig7}
in the case of the 10-state random clock model. In the limit $L_{\rm min}
\rightarrow +\infty$, the critical exponents are estimated to be
$\nu\simeq 1.24(2)$ in the 2D case and $0.99(1)$ in 3D. These values
are in agreement with those obtained for the Potts model:
$\nu\simeq 1.25(3)$ in the 2D case and $0.985(10)$ in 3D~\cite{Anfray}.

\begin{figure}
    \centering
    \includegraphics[width=0.50\textwidth]{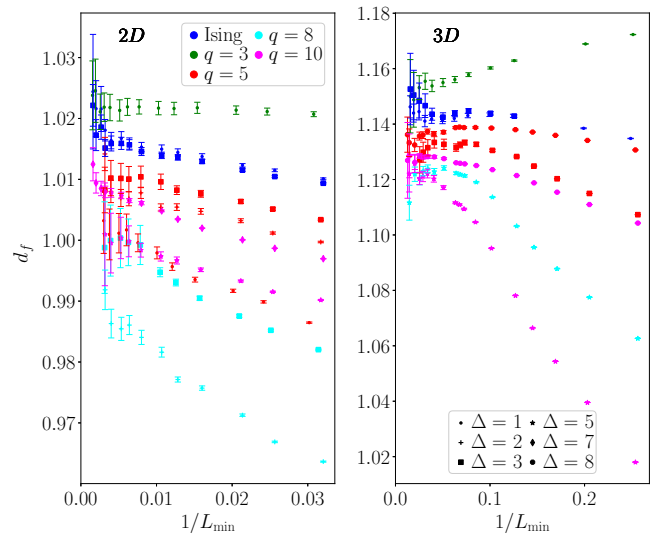}
    \caption{Magnetic fractal dimension $d_f$ extracted from the finite-size
    behavior of the average magnetic moment $\bar\mu$ of the last cluster
    versus the inverse $1/L_{\rm min}$ of the smallest lattice size $L_{\rm min}$
    taken into account in the fit.
    The left figure corresponds to the 2D random $q$-state clock model
    and the right one to the 3D model. The color of the symbols is related to the
    number of states $q$ and their shape to the strength of the initial disorder.}
    \label{fig8}
\end{figure}

The average magnetic moment of the last macro-spin at the end of the RG process
is expected to scale as
    \begin{equation}
    \bar\mu\sim L^{d_f}
    \label{FSS-df}
    \end{equation}
Effective fractal dimensions $d_f(L_{\rm min})$ are computed by a simple
power-law fit. The estimates are presented on Fig.~\ref{fig8}
versus the smallest lattice size $L_{\rm min}$ considered in the fit.
In the 2D case, the effective exponents for different numbers of states $q$ get
closer when $L_{\rm min}$ is increased but they are still incompatible in the
limit $L_{\rm min}\rightarrow +\infty$
when considering the error bars. We note however that, when the disorder strength
$\Delta$ is increased, the effective fractal dimensions $d_f(L_{\rm min})$ of the
Ising model $(q=2)$ are relatively stable. The effective fractal dimensions for
$q=8,10$ and $5$ are systematically lower than the Ising values but the difference
is much smaller at strong disorder than at weak disorder. It is therefore plausible
that the fractal dimensions eventually become compatible for all $q$-state clock
models at stronger disorders. The Ising estimates being the more stable, the
fractal dimension would be $d_f\simeq 1.018$. Note that it was shown that
the fractal dimensions of the 2D $q$-state random Potts model is compatible
with this value ($1.021(5)$) for all number of states $q$~\cite{Anfray}.
A strong dependence of the scaling
corrections on the disorder strength  $\Delta$ is also observed in the 3D case.
The effect is particularly important for the $q=10$ clock model: the slope of
the effective fractal dimensions $d_f(L_{\rm min})$ depends strongly on the
disorder strength. Nevertheless, for the strongest disorder $\Delta=8$, the
fractal dimensions of the $q=5$ and $10$-clock models are compatible within
errors bars with the value $d_f\simeq 1.132(6)$. The latter is close, although
not compatible within error bars, with the estimate $1.155(8)$ of the fractal
dimension of the 3D random $q$-state Potts model~\cite{Anfray}.

\begin{figure}
    \centering
    \includegraphics[width=0.50\textwidth]{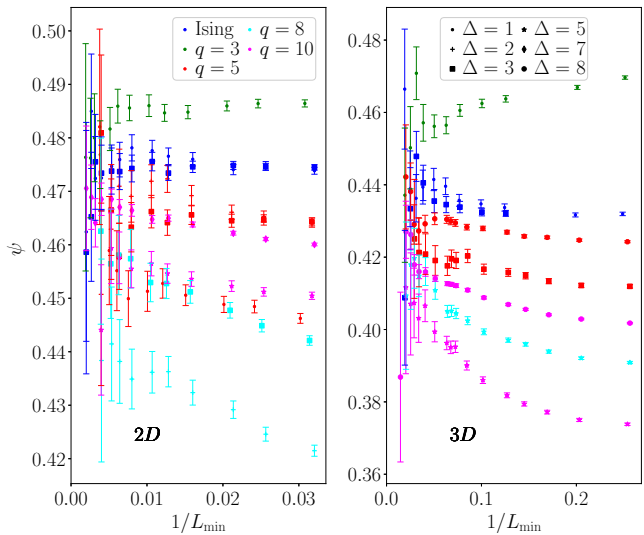}
    \caption{Critical exponent $\psi$ extracted from the finite-size behavior
    of the logarithm of the energy gap $\overline{\ln\Delta E}$ of the last cluster
    versus the inverse $1/L_{\rm min}$ of the smallest lattice size $L_{\rm min}$
    taken into account in the fit.
    The left figure corresponds to the 2D random $q$-state clock model
    and the right one to the 3D model. The color of the symbols is related to the
    number of states $q$ and their shape to the strength of the initial disorder.}
    \label{fig9}
\end{figure}

The dynamical exponent being infinite at the IDFP, the energy gap scales with
the lattice size as
    \begin{equation}
    \Delta E\sim a\ \!e^{-b\ \!L^\psi}
    \label{FSS-Psi}
    \end{equation}
where $a$ and $b$ are two constants. An effective exponent $\psi(L_{\rm min})$
is estimated by a non-linear fit of the average $\overline{\ln\Delta E}$ of
the logarithm of the energy gap as $\ln a+b\ \!L^\psi$. The estimates of
this exponent are presented on Fig.~\ref{fig9}. The error bars are much larger
than for the fractal dimensions $d_f$ and the exponents $\nu$. Again, the
scaling corrections are observed to depend both on the number of states $q$
and on the disorder strength $\Delta$. In the 2D case, when considering
only the disorder strengths $\Delta\ge 3$, the effective critical exponents
in the limit $L_{\rm min}\rightarrow +\infty$ are compatible within error bars
with the value $\psi\simeq 0.467(10)$. They are also compatible with the
estimate $0.48(2)$ obtained for the 2D random quantum Potts model~\cite{Anfray}.
In the 3D case, the error bars are even
larger. For disorder strength $\Delta\ge 3$, the critical exponents are
compatible with $\psi\simeq 0.43(2)$. For the strongest disorder considered
$\Delta=8$, the $q=5$ and 10 clock models are compatible with the value
$\psi\simeq 0.425(5)$. For the 3D random quantum Potts model, the estimate
was $0.46(4)$~\cite{Anfray}.

\begin{acknowledgments}
The numerical simulations of this work were performed at the meso-center
eXplor of the universit\'e de Lorraine under the project 2018M4XXX0118.
\end{acknowledgments}

\section{Conclusions}
We have analyzed the phase transition undergone by the 2D and 3D random quantum
clock and Potts models. By studying the analogue of the classical McCoy-Wu
model, a single phase transition is observed as a peak of the magnetic
susceptibility. It is however not possible with our data to decide whether
the susceptibility diverges only at one critical temperature or over a finite
range of temperatures as expected in Griffiths phases. The lattice sizes
for which an exact average over disorder can be performed are too small.
We then analyzed the integrated probability distribution of the logarithm
of the local susceptibilities which is expected to decay algebraically in
Griffiths phases~\cite{Rieger}. No such decay could be observed for the
clock model. In contrast, an algebraic decay is clearly observed over a
wide range of several decades for the Potts model. However, close to the
critical point, this range is reduced by strong Finite-Size effects.
These effects could be interpreted in light of the physical mechanism
proposed in Ref.~\cite{Rieger}. Nevertheless, it is not clear to us
why the Potts model display Griffiths phase whereas the clock model
does not. A possible explanation of the absence of Griffiths
phase could be that $\ell_{\rm min}>L_\perp$ for the clock model,
i.e. a simple Finite-Size effect. It is also possible that Griffiths
phases would appear at stronger disorder.
\\

In the second section of this paper, we studied the critical behavior
of the 2D and 3D random quantum clock models by Strong Disorder Renormalization
Group (SDRG) and compared the estimated exponents with the known values for
the Ising and Potts models. We have shown that the use of second-order
perturbation theory in the determination of the SDRG rules are responsible
for a chaotic behavior of the RG flow at weak disorder. Nevertheless,
at strong enough disorder, our estimates of the critical exponents
are compatible, or at least close in some cases, to the Ising and Potts
values leading to the conclusion that the super-universality class of the
random Ising and Potts models encompasses the clock model for any number
of states.

\end{document}